# Interplay of superconductivity and ferromagnetism in ferromagnetic-semiconductor-based Josephson junctions

Hirotaka Hara[1,*], Lukas Baker[2,*], Axel Leblanc[2], Shingen Miura[1], Keita Ishihara[1], Melissa Mikalsen[2], Patrick J. Strohbeen[2], Jacob Issokson[2], Masaaki Tanaka[1,3,4], Javad Shabani[2,†], and Le Duc Anh[1,3,†]

[1] *Department of Electrical Engineering and Information Systems, The University of Tokyo, Tokyo, Japan*

[2] *Center for Quantum Information Physics, New York University, New York, USA*

[3] *Center for Spintronics Research Network, The University of Tokyo, Tokyo, Japan*

[4] *Institute for Nano Quantum Information Electronics, The University of Tokyo, Tokyo, Japan*

\* These authors contribute equally to this work.

†Corresponding author: anh@cryst.t.u-tokyo.ac.jp, jshabani@nyu.edu

**Abstract**

The interplay between superconductivity and ferromagnetism has long been pursued as a route to unconventional Josephson effects, yet suitable material platforms remain limited. Here we report Josephson junctions based on epitaxial Al/InAs/(Ga,Fe)Sb heterostructures grown by low-temperature molecular beam epitaxy, achieving atomically abrupt superconductor / semiconductor / ferromagnetic-semiconductor interfaces. The devices exhibit clear proximity-induced superconductivity, including multiple Andreev reflections and gate-tunable supercurrents, confirming transparent coupling across the hybrid structure. Under perpendicular magnetic fields, the junctions reveal highly unconventional Fraunhofer interference patterns with hysteresis, flux jumps, asymmetric lobe evolution, and clear nonreciprocity, providing strong evidence of induced ferromagnetism and broken time-reversal symmetry, and possible existence of the edge channels in the Josephson junctions. Our results demonstrate that ferromagnetic semiconductor heterostructures can serve as a highly tunable platform for exploring proximity-induced superconductivity and superconducting diode effects, and for advancing device concepts at the intersection of magnetism and quantum electronics.

**Introduction**

The coexistence of superconductivity and ferromagnetism has been actively investigated for several decades, as their antagonistic nature gives rise to intriguing phenomena such as Josephson π-junctions [1], spin-triplet superconductivity [2], and superconducting diode effect [3]. To realise π-junctions, ferromagnetic barriers are often introduced between two superconductors using diluted ferromagnetic metals [4][5]. In contrast, non-diluted ferromagnets, with their large exchange energies, drastically suppress the superconducting coherence length in the ferromagnetic region, where only spin-triplet supercurrents can survive if present. Such Josephson junctions incorporating ferromagnets are valuable not only for elucidating the interplay between superconductivity and ferromagnetism but also for developing superconducting device applications [6]-[8]. However, their physical properties are highly sensitive to the thickness and composition of the ferromagnetic materials, which presents challenges for practical devices [5][9].

An alternative approach to realizing π-junctions employs superconductor/semiconductor heterostructures. In semiconductors with a large g-factor, large Zeeman splitting can be induced by an external magnetic field, resulting in non-zero pairing momentum of Cooper pairs and the emergence of the Fulde–Ferrell–Larkin–Ovchinnikov (FFLO) state [10][11]. This state also leads to π-phase shifts in Josephson junctions based on semiconductors [12]-[15]. Although the use of external magnetic fields are generally undesirable for scalable architectures, superconductor/semiconductor heterostructures have been widely adopted as a platform for superconducting quantum devices, owing to the high controllability of proximity-induced superconductivity by applying an electrical gate voltage [16]. In particular, epitaxially grown heterostructures with atomically clean interfaces are promising for applications such as gatemons [17][18] and topological qubits [19][20].

In pursuit of a versatile platform for both fundamental physics and device applications, this study focuses on Fe-doped narrow-gap III-V ferromagnetic semiconductors (FMSs) as a host for proximity-induced superconductivity. Narrow-gap III–V FMSs are particularly promising candidates because they are expected to form Ohmic and transparent interfaces with superconductors, reducing interfacial energy barriers, as demonstrated in Al/InAs heterostructures [21]-[24]. Moreover, both carrier concentration and ferromagnetic exchange energy in Fe-doped narrow-gap III–V FMSs can be tuned by applying an electrical gate voltage [25][26]. These experimental achievements suggest that superconductor/Fe-doped III–V FMS heterostructures provide a unique platform for systematically studying proximity-induced superconductivity in ferromagnets, in line with theoretical predictions [27][28]. Additionally, a semiconductor system exhibiting both spontaneous spin splitting and proximity-induced superconductivity may play a key role in the context of Majorana physics.

To date, proximity-induced superconductivity in FMSs has only been demonstrated in Nb/Fe-doped InAs ((In,Fe)As) heterostructures [29][30]. In that study, long-range proximity effects and hysteretic behaviour were observed in superconducting (In,Fe)As channels through transport measurements, suggesting the existence of spin-triplet supercurrents. While these results highlight the importance of further studies on superconductor/FMS heterostructures, several limitations remain. First, the Nb/(In,Fe)As heterostructures were grown *ex-situ* (after growing InFeAs by molecular beam epitaxy (MBE), the sample was exposed to air and then Nb was deposited in another vacuum chamber), which may compromise interface quality. Second, gate voltage control of the (In,Fe)As properties is challenging and has only been realised using electronic double-layer transistor (EDLT) methods with liquid electrolytes [25][26].

To overcome these limitations, we focus here on Al/InAs/(Ga,Fe)Sb heterostructures, where the whole heterostructures are grown *in-situ* by MBE in an ultrahigh vacuum without exposure to air, ensuring clean interfaces. In InAs/(Ga,Fe)Sb heterostructures, the (Ga,Fe)Sb layer becomes insulating at low temperatures [31], so that electrical transport primarily occurs through the InAs layer [32]. Importantly, the InAs channel exhibits large and tunable spin splitting in the conduction band resulting from a strong magnetic proximity effect by the adjacent FMS (Ga,Fe)Sb, owing to large penetration of the electron wavefunction in InAs to the (Ga,Fe)Sb layer in the type-III band alignment [32][33]. Spin splitting as large as 18 meV in InAs has been observed, which is controllable by modulation of the carrier wavefunction and concentration via gate voltage [33]. Therefore, superconductor Al / InAs / Fe-doped FMS (Ga,Fe)Sb heterostructures represent a promising platform for investigating proximity-induced superconductivity and ferromagnetism under electrostatic control. This central concept of this study is shown in Fig. 1.

**Results**

The Al/InAs/(Ga,Fe)Sb heterostructure grown by low-temperature molecular beam epitaxy (LT-MBE) is shown in Fig. 2(a). After growing AlSb/AlAs buffer layers on a GaAs (001) semi-insulating substrate, a 5 nm-thick (In,Ga)As / 15 nm-thick InAs / 20 nm-thick (Ga,Fe)Sb heterostructure, from top to bottom, was grown at 250°C to suppress Fe segregation in the (Ga,Fe)Sb layer. The thin (In,Ga)As layer containing 19% indium (In) is inserted between the Al and the InAs layers to make better superconductor/semiconductor interfaces. The (Ga,Fe)Sb layer containing 14.6% iron (Fe) is an FMS layer to induce a magnetic proximity effect in the adjacent InAs layer. A 10 nm-thick Al film was subsequently grown *in situ* at a temperature below −40°C. Transmission electron microscopy (TEM) images in Figs. 2(b) and 2(c) show that the semiconductor layers retain the zinc-blende-type crystal structure throughout MBE growth. Figure 2(c) further indicates that the Al film grows along the [111] direction of the face-centered cubic crystal, consistent with the X-ray diffraction results presented in Supplementary Material. The TEM analysis also confirms that the Al/(In,Ga)As interface is clean and atomically sharp.

We first characterise the sample transport and magnetic properties. Magnetotransport of the InAs layer at 3.7 K under a perpendicular magnetic field was measured after removing the Al layer by wet etching. As shown in Fig. 2(d), we observe clear Shubnikov-de Haas (SdH) oscillations, from which we estimated the electron mobility $\mu$ to be $3.2\times10^3$ cm²V⁻¹s⁻¹ along the [$\bar{1}$10] crystal direction and $1.8\times10^3$ cm²V⁻¹s⁻¹ along the [110] direction, and the electron carrier density $n$ to be $3.5\times10^{12}$ cm⁻². These mobility values are the highest reported so far for two-dimensional electrons in InAs/(Ga,Fe)Sb heterostructures [32]. Figure 2(e) presents magnetic circular dichroism (MCD) and superconducting quantum interference device (SQUID) magnetometry data. At 5 K, both measurements reveal clear hysteresis loops in good agreement with each other, which gradually vanish with increasing temperature. The Curie temperature is estimated to be 25 K from an Arrott plot of the MCD–$H$ data. These results demonstrate that high-quality Al(111)/InAs/(Ga,Fe)Sb heterostructures with atomically sharp interfaces are successfully realised by using LT-MBE.

Lateral Josephson junctions are then fabricated from this heterostructure using electron-beam lithography and wet etching (see Methods). A schematic of a gated junction and a scanning electron microscopy image of a real device are shown in Fig. 3(a) and (b), respectively. The junctions reported here are aligned with the [110] direction of the substrates, along which the corresponding mean free path $l_e$ is estimated to be 57 nm from electron concentration $n$ and mobility $\mu$. The channel length $L$ is 100 nm and the width is 5 μm, placing the device in a diffusive transport regime.

Figure 3(c) displays the *V–I* characteristics of an ungated junction at 30 mK. Unless otherwise stated, the *V–I* data were acquired by sweeping from zero to high bias currents in both plus and minus directions. The junction exhibits a zero-resistance state at low bias currents and nearly linear behaviour at higher currents. From these data, we extracted a critical current $I_c$ = 3.00 μA, an excess current $I_{ex}$ = 1.91 μA, and a normal resistance $R_N$ = 52.5 Ω. Figure 3(d) shows the corresponding d$I$/d$V$ spectrum, which exhibits several peaks resulting from multiple Andreev reflections (MARs) [34]. These peaks satisfy the relation $V_n = 2\Delta/ne$, where $\Delta$ is the superconducting gap of Al and $n$ is an integer constant, and are observable up to $n = \pm 6$. A linear fit of $V_n$ versus $1/n$ (Fig. 3(e)) yields a superconducting energy gap $\Delta = 171$ μeV, which agrees well with that estimated from the critical temperature ($\Delta_0 = 1.76 k_B T_c = 196$ μeV). The observation of MARs up to sixth order indicates an anomalously long inelastic scattering length in InAs (> 600 nm), although $l_e$ (= 57 nm) is less than the junction length $L = 100$ nm. As shown in Fig. 3(f), even under applying a gate voltage $V_g$ (< 0 V), the MAR peaks persist, while the supercurrent decreases systematically with increasing $|V_g|$ (negative $V_g$ depletes electrons in the InAs channel). This demonstrates successful modulation of the critical current by electrostatic gating, which is important for practical device applications and provides further evidence for proximity-induced supercurrent through the semiconductor InAs layer.

The Josephson junction interface transparency can be quantitatively evaluated by parameters $\alpha = eI_c R_N/\Delta = 0.92$ and $\alpha' = eI_{ex} R_N/\Delta = 0.59$. The $\alpha$ value corresponds to 29%/44% of the ballistic/diffusive theoretical limits, and the $\alpha'$ value corresponds to 22%/40% of the ballistic/diffusive theoretical limits, respectively [35]-[38]. While these data are not the highest among reported values, they exceed those of superconductor/semiconductor junctions prepared by sputtering and are comparable to those of epitaxially grown heterostructures [39]. These results confirm proximity-induced superconductivity with relatively high interface transparency in our Al/Fe-doped FMS heterostructures.

Next, we examined electrical transport under external magnetic fields in ungated (Fig. 4) and gated (Fig. 5) Josephson junction devices. Figures 4(a) and 4(b) show differential resistance as a function of perpendicular magnetic field $B$ and bias current $I$ when $B$ is swept in opposite directions. Black arrows indicate the magnetic field sweeping directions, and $B_{ini}$ denotes the initialisation field applied to align the (Ga,Fe)Sb magnetisation before each measurement. The Fraunhofer patterns, where the superconducting regions are shown in yellow, exhibit unique and unconventional features that are distinctly different from the $B$-symmetric patterns in non-magnetic Josephson junctions [40][41]. In particular, the critical current $I_c$ display maxima at non-zero $B$, which have the same magnitude but opposite sign depending on the magnetic-field sweeping direction. Such hysteresis behaviour provides strong evidence of induced ferromagnetism in the superconducting InAs channel. Moreover, there are several noteworthy features: Unequal oscillation periods of the critical current between successive lobes; and non-monotonic decay of the $I_c$ away from the central lobe, and numerous discontinuous points during magnetic-field sweeps, which we identify as flux jumps (the flux jumps, however, do not occur at exactly the same magnetic field across repeated measurements). These deviations from an ideal Fraunhofer pattern may arise from non-uniform magnetic-field and/or current distributions in the junction [42]-[44]. Such nonuniformity may originate from the magnetisation reversal of magnetic domains in (Ga,Fe)Sb, which affects the magnetic-field distribution, and from accumulation of electrons at the two edges of the InAs channel, which will be discussed later.

We conducted a fine-field scan around the maximum critical current (~ −0.5 mT) in Fig. 4(a), where $I_c$ oscillations are observed without flux jumps, as shown in Fig. 4(c). The horizontal axis represents the magnetic field relative to the center field $B_{shift}$ (~ −0.5 mT) defined as the point of maximal $I_c$. The oscillation period of 0.14 mT, which is determined with the fitting curve expressed by the conventional Fraunhofer equation $I_c(B) = I_0 \text{sinc}(\pi \Gamma B S/\Phi_0)$ where $\Gamma$, $S$, $\Phi_0$ is a flux focusing factor, the junction size, and the flux quantum, is much smaller than the theoretical value of 0.86 mT estimated from the junction size. This indicates the flux focusing factor $\Gamma$ of 6.2, likely arising from the Meissner effect of Al electrodes and stray fields from (Ga,Fe)Sb. Furthermore, the pattern is slightly skewed and the critical current exhibits pronounced nonreciprocity, with a diode efficiency $\eta = (|I_c^+| - |I_c^-|)/(|I_c^+| + |I_c^-|)$ reaching 20% at the minimum of the Fraunhofer pattern (violet circles in Fig. 4(c)). Here, the superscripts ± denote opposite current directions. $\eta$ is nearly antisymmetric with respect to the relative field $B^* = B - B_{shift}$. The relations $I_c^+(+B^*) \neq I_c^-(+B^*)$ and $I_c^+(+B^*) \neq I_c^+(-B^*)$ manifest the broken time-reversal symmetry in the junction, providing further evidence of proximity-induced magnetism in the superconducting InAs channel. Combined with space-inversion-symmetry breaking inherent to the asymmetric heterostructure stacking, such time-reversal-symmetry breaking possibly induces the observed superconducting diode effect.

To gain a further insight into the junction characteristics, fast Fourier transformation (FFT) was performed on the Fraunhofer patterns shown in Fig. 4(c). The FFT analysis provides a spatial mapping of the current distribution along the in-plane $y$-direction, which is perpendicular to the current flow, as illustrated in Figs. 4(d) and 4(e). The details of the FFT procedure are described in Supplementary Material and in a reference [45]. The fitted current profile does not appear completely flat due to the limited number of lobes in the Fraunhofer pattern shown in Fig. 4(c). The obtained current distribution indicates a tendency for the supercurrent to concentrate near the physical edges of the junction, in contrast to the more homogenous profile derived from the fitting curve expressed by a sinc function.

As shown in Fig. 5(a)-(d), Fraunhofer patterns at gate voltages of $V_g = 0$ V and

−10 V under opposite sweeping directions of magnetic field consistently exhibit hysteresis behaviour depending on the sweeping direction; the overall shapes remain similar while the critical current decreases with applying gate voltage $V_g$. Figure 5(e) shows a colored map of the critical current $I_c$ versus the perpendicular magnetic field, which has the same set up as those in Fig. 5(b)(d), when $V_g$ is further swept from −10 V to −25 V. Two distinct regions with relatively large $I_c$ are observed; peak 1 (from −1.6 to −0.7 mT) and peak 2 (from −0.6 to 0 mT). As shown in Fig. 5(f), upon decreasing $V_g$ the decay of $I_c$ in peak 1 closely follows the decrease of the electron carrier density $n$ in the InAs channel, which was obtained by Hall measurements at 5 K. This indicates that the change of $I_c$ in peak 1 directly results from the depletion of Cooper pairs in the junction channel. On the other hand, the decay of $I_c$ upon decreasing $V_g$ in peak 2 is noticeably weaker than either that of peak 1 or the carrier density. This unusual behaviour suggests that superconducting transport in the junction involves multiple conduction channels, rather than a single uniform one, which is likely consistent with current density mapping obtained by FFT (Fig. 4(d)).

Furthermore, there are also pieces of evidence supporting edge channels in the Fraunhofer patterns upon closer inspection. First, the Fraunhofer patterns exhibit node-lifting, i.e., a nonzero critical current at the local minima of the pattern, as clearly observed in the first node (the nearest nodes from the maximum critical current) but not in the second nodes (the second nearest nodes from the maximum one) of the data in Fig. 4(c). In conventional Fraunhofer interference, these minima should be zero. Previously, observations of node-lifting in the Fraunhofer patterns were attributed to the existence of edge channels, which are either topologically non-trivial [46]-[49] or trivial [50]. Second, the maximum critical current in each lobe does not decay monotonically with the distance from the central lobe but instead alternates between large and small values—a phenomenon known as even–odd modulation. This can be clearly observed around 0.5 mT in Fig. 4(a) and −0.5 mT in Fig. 4(b). The even−odd modulation occurs by the beating between $h/2e$ ($2e$-transport) and $h/e$-periodicity ($e$-transport) of the Fraunhofer patterns [51]. While the $h/2e$ periodicity is conventional, the $h/e$-periodicity is often associated with crossed Andreev reflections involving two edge channels [50][52]-[55]. In our junctions, however, this is opened to discussion considering that the estimated superconducting coherent length (~1.7 μm) is much shorter than the junction width (~5 μm). Nevertheless, these observations collectively point to the possible existence of superconducting edge channels, carrying both Cooper pairs and quasi-particles, in the InAs region.

Edge channels have indeed been reported at the two sides of InAs/(Ga,Fe)Sb heterostructures [56]. Under a perpendicular magnetic field, their normal conductance changes with opposite sign, producing large odd-parity magnetoresistance attributed to simultaneous breaking of time-reversal symmetry (via the magnetic proximity effect) and inversion symmetry at the edges. Consistent with this picture, the FFT analysis in Fig. 4(d) reveals a slight asymmetry between the supercurrent densities at the two edges. The skewness of the Fraunhofer pattern may involve the asymmetric contributions from the edge channels, as suggested previously in twisted bilayer graphene Josephson junctions [57]. Further study is needed to fully understand the unconventional interference behaviour in these ferromagnetic junctions.

In conclusion, we have observed superconducting proximity effects in Josephson junctions fabricated from MBE-grown Al/InAs/FMS (Ga,Fe)Sb heterostructures. The junctions exhibit gate-tunable critical currents, a key feature for device applications exploiting semiconductor properties. Furthermore, they display Fraunhofer patterns with magnetic hysteresis, indicative of induced ferromagnetism in the superconducting InAs channel. Transport characteristics, including Fraunhofer interference, also provide possible evidence of edge-channel superconducting transport. Previously, a spin splitting energy of up to 18 meV has been observed in InAs when interfaced with (Ga,Fe)Sb containing 20% Fe and a Curie temperature of 300 K [33]. This spin-splitting energy is large enough to support a $\pi$-phase transition, as it significantly exceeds the Zeeman energy in semiconductors [12]-[15] and is comparable to the exchange energy in diluted ferromagnets [58]-[60]. However, the (Ga,Fe)Sb used in this study has a Curie temperature of only 25 K and small remanent magnetisation (see Supplementary Material), resulting in a smaller spin-splitting energy that may be insufficient for reversing the phase of the superconducting wavefunction across the junction length, though it could help reduce spin-related scattering and enhance InAs mobility. Therefore, Al–InAs/(Ga,Fe)Sb–Al heterostructures with improved magnetic properties are required to realise $\pi$-phase transitions in the future work. These gate-controllable ferromagnetic Josephson junctions offer a promising platform for studying the coexistence of superconductivity and ferromagnetism and for developing new superconducting quantum devices.

**Methods**
**Sample preparation**
Al/InAs/FMS (Ga,Fe)Sb heterostructures were grown on GaAs (001) semi-insulating substrates using an MBE system equipped with two connected ultrahigh-vacuum chambers. The native oxide layer was removed in a chamber specialised for III-V semiconductors by heating the substrates to 580°C. Subsequently, GaAs and 5-nm AlAs buffer layers were grown at 570°C, followed by an AlSb buffer layer deposited at 470°C. After the buffer layer growth, the substrate temperature was lowered to 250 °C, and LT-MBE was employed to grow 20-nm $(Ga_{0.854},Fe_{0.146})Sb$, 15-nm InAs, and 5-nm $(In_{0.81},Ga_{0.19})As$ layers. The Fe concentration in (Ga,Fe)Sb was calibrated by secondary ion mass spectrometry (SIMS) and Rutherford backscattering spectroscopy (RBS). Then, the sample was transferred *in-situ* to the other MBE chamber and cooled down by thermal contact to a liquid nitrogen dumper for more than 90 minutes to reach below −40°C. Once sufficiently cooled, a 10-nm-thick Al layer was grown on top of the semiconductor layers.

**Fabrication of Josephson junctions**

Lateral Josephson junctions (JJ) with an Al–InAs/(Ga,Fe)Sb–Al conduction path were fabricated using a top-down process. First, Al electrodes and bonding pads were patterned by electron-beam lithography with PMMA resist, followed by wet etching using Transene Type-D etchant for the Al layer and a mixed solution of phosphoric acid, hydrogen peroxide, and water for the III–V semiconductor layers, including the FMS layer. Trench structures defining the junctions were then formed by electron-beam lithography and etched with Transene Type-D etchant. To fabricate JJs with gate structures, a 40-nm $Al_2O_3$ gate dielectric was deposited by atomic layer deposition at a substrate temperature of 40°C. Subsequently, a gate electrode consisting of 10-nm-thick Ti and 70-nm-thick Au was deposited by electron-beam evaporation, and the gate area was defined by a lift-off process.

**Electrical measurements**

For the evaluation of the lateral JJs, measurements were performed in a low-temperature environment at 30 mK using an Oxford Triton dilution refrigerator. The junctions were characterised under current-biased conditions using either a DC measurement setup (Keithley 6500) or a low-frequency AC lock-in technique (Stanford SR860).

To investigate the influence of the (Ga,Fe)Sb magnetisation on the junction properties, special care was taken during the application of external magnetic fields. Prior to each magnetic-field sweep, an initial magnetic field was applied to align the magnetisation direction of the (Ga,Fe)Sb layer, followed by conventional procedures used in magnetic characterisations. This initialisation ensures that the subsequent measurements can be compared on a consistent basis, even though the process may allow vortices to enter the Josephson junctions.

**Data availability**

The main data that support the findings of this study are available in this article and its Supplementary Information. Additional data are available from the corresponding author upon request.

**Acknowledgments**
This work was supported in part by the Grants-in-Aid for Scientific Research (19K21961, 20H05650, 22K18293, 23K17324, 24H00018, and 25H00840), the CREST program (JPMJCR1777), NEXUS program of JST, and the Spintronics Research Network of Japan (Spin-RNJ). HH thanks financial support from Hirose Foundation, Marubun Research Promotion Foundation, and Overseas Training Program of the School of Engineering, the University of Tokyo, which enabled his visit to and experiments at the Shabani Laboratory at New York University.

**Author contributions**
H.H. and L.B. conceptualized the study, led the experiments, analyzed the data, and prepared the initial draft. A.L., S.M., K.I., M.M., P.J.S., and J.I. grew the samples, fabricated the devices, performed low-temperature measurements, and analyzed the data. M.T., J.S., and L.D.A. supervised the experiments, contributed to data discussions, and revised the manuscript. All authors discussed the results and provided comments on the manuscript.

**Competing interests**
The authors declare no competing interests.

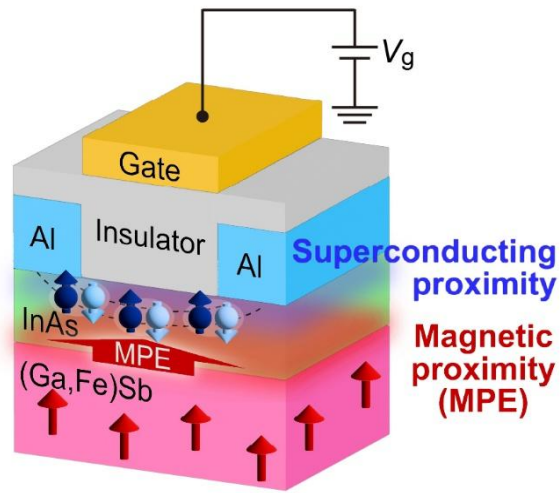

**Figure 1.** Conceptual illustration of this study. Proximity-induced superconductivity and ferromagnetism are introduced into the InAs semiconductor layer from the adjacent overgrown superconducting Al electrodes and underlying Fe-doped FMS (Ga,Fe)Sb layer, respectively. Owing to the semiconductor nature of InAs, both the proximitised superconductivity and the magnetic proximity effect are controlled by applying a gate voltage.

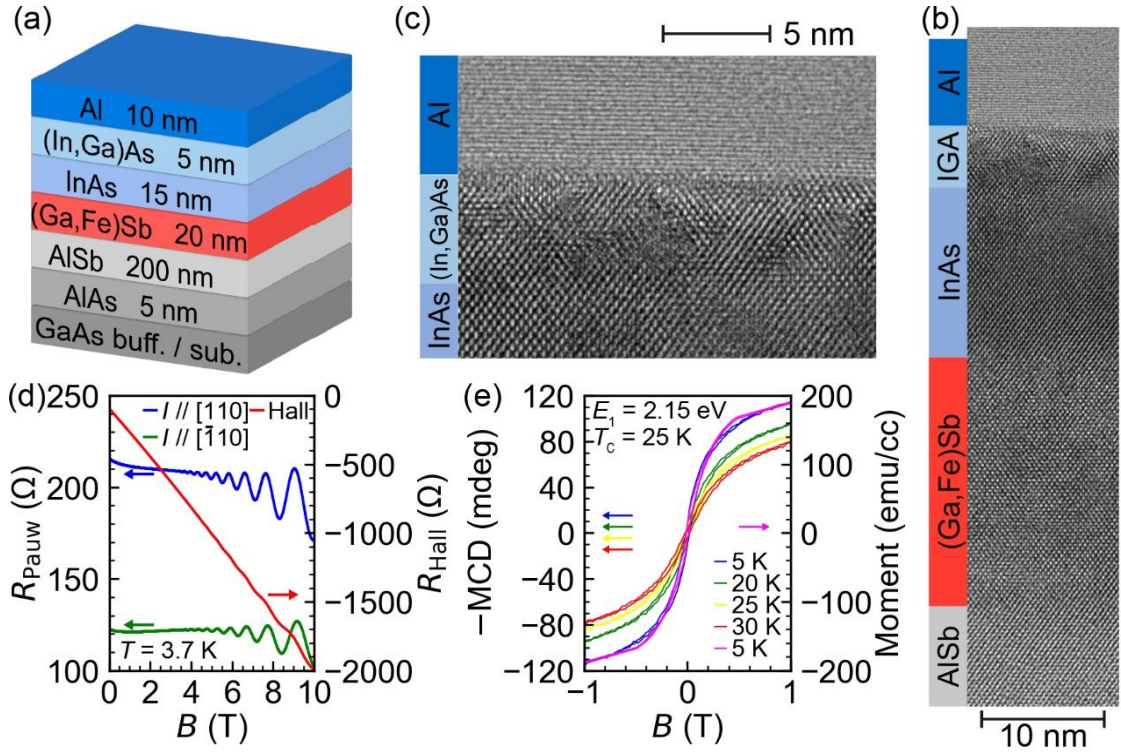

**Figure 2.** (a) Schematic structure of the sample used in this study. (b), (c) Cross-sectional transmission electron microscopy lattice images of the Al / (In,Ga)As / InAs / FMS (Ga,Fe)Sb heterostructure with the incident direction along the [110] axis of the GaAs (001) substrate. The semiconductor layers are grown in the (001) orientation with a zinc-blende-type crystal structure, while the Al layer is grown in the (111) orientation with a face-centered cubic crystal structure. (d) Magnetotransport of the InAs channel under a perpendicular magnetic field at 3.7 K. The resistance of InAs was measured by the Van der Pauw method using a cleaved 5 mm × 5 mm piece. Blue and green curves represent the resistance with current applied along the [110] and [$\bar{1}$10] crystal axes, respectively, while the red curve shows the Hall resistance. (e) Magnetic properties of the (Ga,Fe)Sb layer obtained by MCD and SQUID measurements. MCD measurements were performed at the critical point energy $E_1$ = 2.15 eV of GaSb. MCD and SQUID results show good agreement at 5 K.

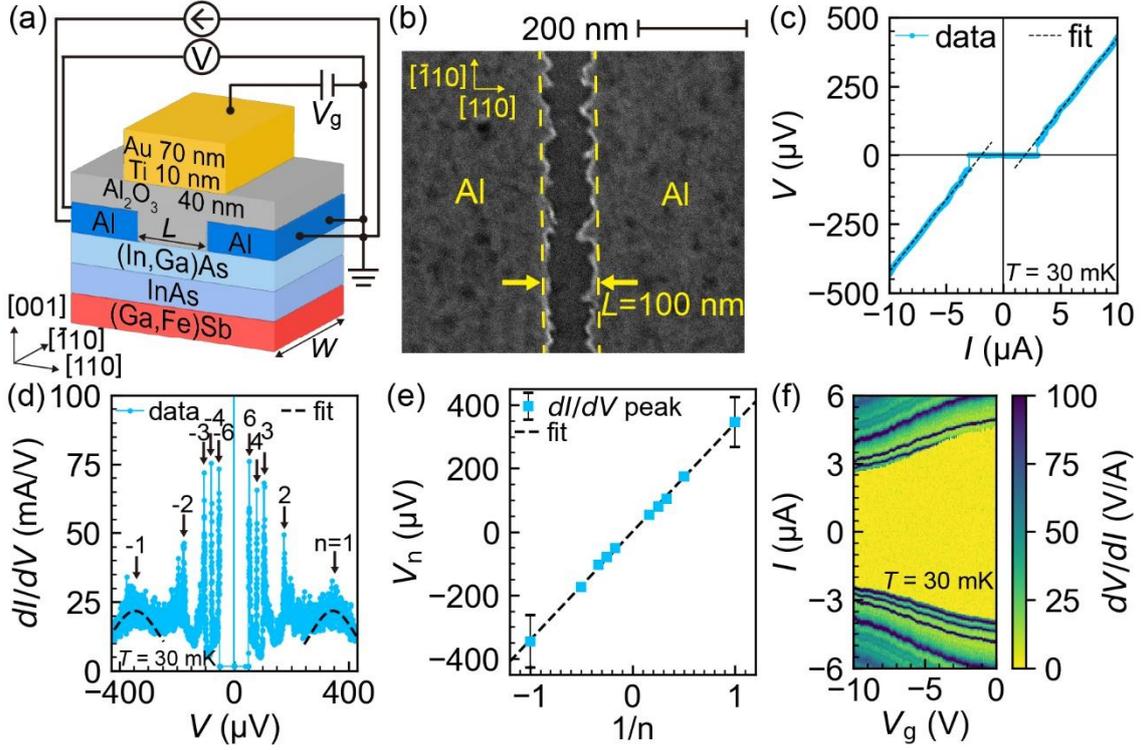

**Figure 3**. (a) Lateral Josephson junction device with a top gate electrode examined in this study. DC measurement setup is also shown. The junction length $L$ and width $W$ are 100 nm and 5 μm, respectively. (b) Scanning electron microscopy image of a junction device without a gate. The junction length is estimated to be ~100 nm at the longest part between the two Al electrodes. (c) Current dependence of the junction voltage ($V – I$ characteristic) in a device without a gate at 30 mK. The dotted line represents interpolation from high-bias data using $V = R_N(I \pm I_{ex})$. (d) Voltage dependence of d$I$/d$V$ calculated from (c). The peak positions $V_n$ correspond to multiple Andreev reflections, following $V_n = 2\Delta/ne$. Peaks at $n = \pm 1$ are relatively broad and were fitted with Gaussian functions to estimate their positions. (e) Relation between the d$I$/d$V$ peak voltages $V_n$ and the inverse reflection order $1/n$, extracted from (d). (f) Dependence of the junction resistance d$V$/d$I$ on the gate voltage $V_g$ and bias current $I$. The current $I$ was swept from negative to positive bias.

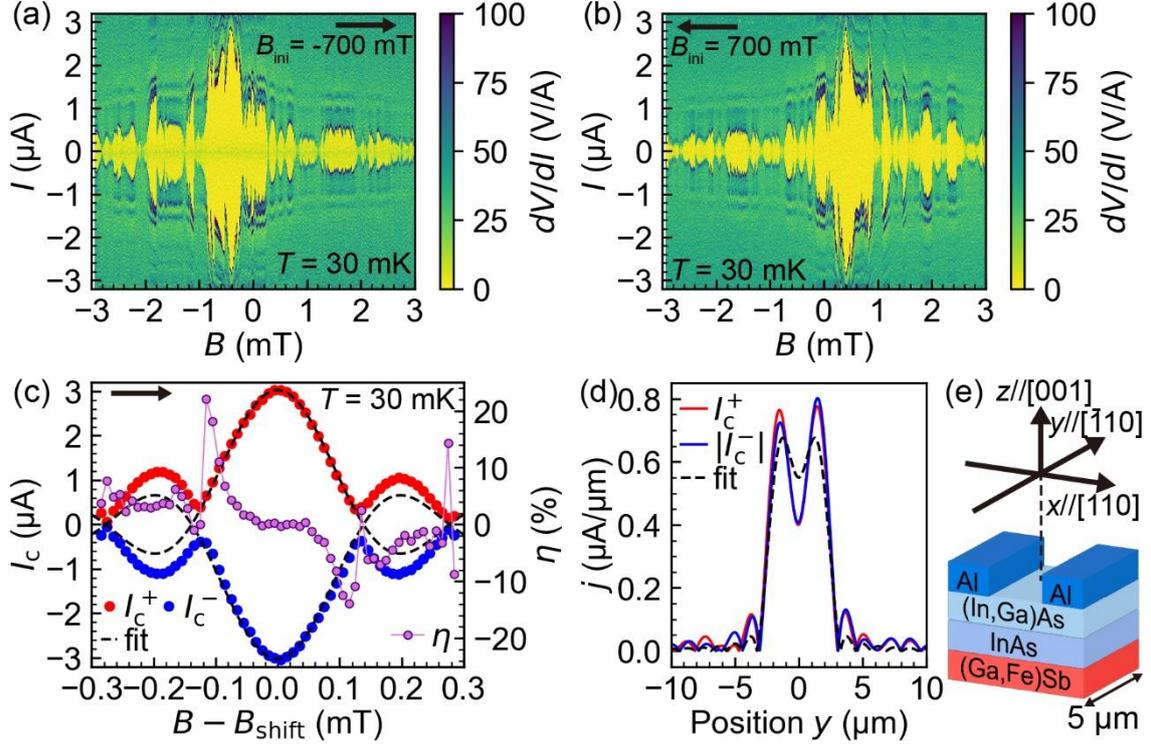

**Figure 4.** (a), (b) Differential resistance of the junction without a gate as a function of magnetic field $B$ and current bias $I$. Black arrows indicate the sweeping directions of $B$, and $B_{ini}$ denotes the initial field applied before sweeping. (c) Fine-field dependence of the critical current $I_c$ and diode efficiency $\eta$. $B_{shift}$ corresponds to the field at which the maximum critical current is observed in the measurement setup of (a). No initial field was applied in this measurement. In all panels, the magnetic field is applied perpendicular to the junction plane. (d) Spatial distribution of the supercurrent density $j$ flowing through the junction obtained by FFT analysis of the data shown in (c). The lateral axis position $y$ (// $[\bar{1}10]$) shown in (e) corresponds to the in-plane direction perpendicular to the flowing current. (e) A schematic Josephson junction image without a gate with the definition of coordinate axes. The origin is set as the center position of the junction.

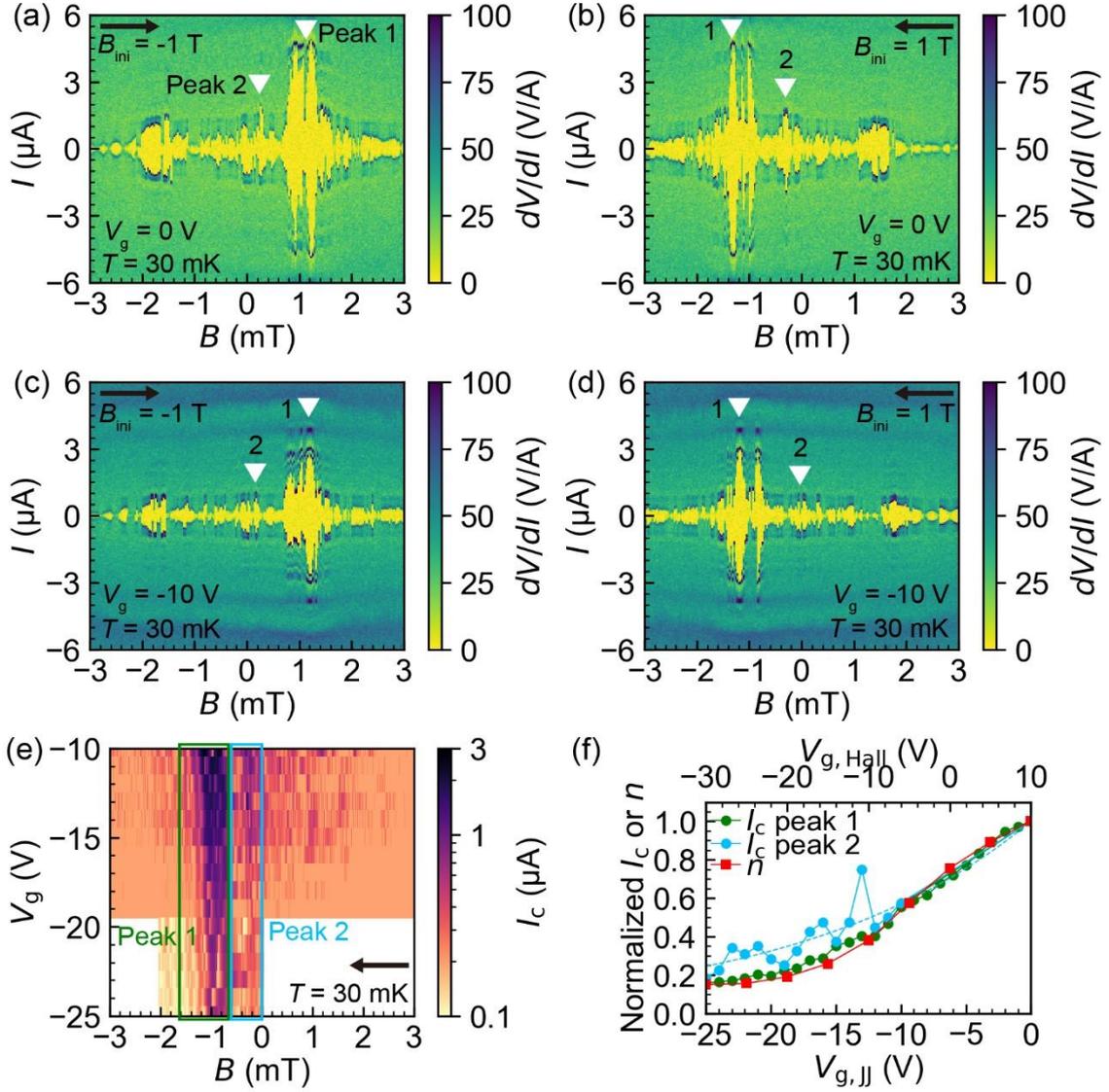

**Figure 5.** (a)-(d) Differential resistance of the gated junction as a function of the magnetic field $B$ and current bias $I$. Black arrows indicate the sweeping directions of $B$, and $B_{ini}$ denotes the initial field applied before sweeping. (a) and (b) were measured at $V_g = 0$ V, while (c) and (d) were measured at $V_g = -10$ V. (e) Critical current $I_c$ as a function of the magnetic field $B$ and gate voltage $V_g$. (f) Dependence of the normalised critical current $I_c$ and carrier concentration $n$ on the gate voltage $V_g$, normalised by their respective saturation values. $I_c$ peak 1 is obtained at peak 1 from the field range −1.6 to −0.7 mT, and $I_c$ peak 2 is obtained at peak 2 from −0.6 to 0 mT in (e). The sky-blue dashed line represents an exponential fit provided as a guide to the eye.

# Supplementary Material

# Interplay of superconductivity and ferromagnetism in ferromagnetic-semiconductor-based Josephson junctions

Hirotaka Hara[1*], Lukas Baker[2*], Axel Leblanc[2], Shingen Miura[1], Keita Ishihara[1], Melissa Mikalsen[2], Patrick J. Strohbeen[2], Jacob Issokson[2], Masaaki Tanaka[1,3,4], Javad Shabani[2,†], and Le Duc Anh[1,3,†]

[1] *Department of Electrical Engineering and Information Systems, The University of Tokyo, Tokyo, Japan*

[2] *Center for Quantum Information Physics, New York University, New York, USA*

[3] *Center for Spintronics Research Network, The University of Tokyo, Tokyo, Japan*

[4] *Institute for Nano Quantum Information Electronics, The University of Tokyo, Tokyo, Japan*

* These authors contribute equally to this work.

†Corresponding author: anh@cryst.t.u-tokyo.ac.jp, jshabani@nyu.edu

**Superconducting Al thin film properties grown on an InAs/(Ga,Fe)Sb heterostructure**

The growth of superconducting Al thin films on ferromagnetic semiconductor (FMS) heterostructures has not yet been well established, therefore, the properties of Al should be carefully examined. Figure S1(a) shows the X-ray diffraction (XRD) pattern of the sample structure presented in Fig. 2(a). Distinct zinc-blende–type GaAs (004) and AlSb (004) peaks are observed at $2\theta$ = 66.07° and 60.12°, respectively. The smaller peak around 61.2° is likely attributable to (Ga,Fe)Sb and InAs. In addition, a face-centered cubic Al (111) peak appears at 38.5°, consistent with the growth orientation identified in the TEM image in Fig. 2(c).

Figures S1(b) and (c) show the electrical transport properties of the Al thin film for evaluation of its superconductivity. Figure S1(b) presents the temperature dependence of the Al resistance, which exhibits a sharp drop at 1.30 K, reaching nearly zero below this temperature. The resistance data shows slight noise, possibly due to measurements during the cooling process. From the transition temperature $T_c$ = 1.30 K, the superconducting gap is estimated as $\Delta_0$ = 1.76 $k_B T_c$ = 196 μeV, which agrees well with the value obtained from multiple Andreev reflections in the main text. Figure S1(c) shows the magnetic field dependence of the Al resistance, measured with an in-plane field aligned along the [$\bar{1}$10] direction of the GaAs (001) substrate. The critical field $B_c$ = 1.16 T is comparable to previously reported values [S1]. These results confirm that the Al layer in our sample maintains suitable superconducting properties for observing the proximity effect, even when grown on the FMS heterostructure.

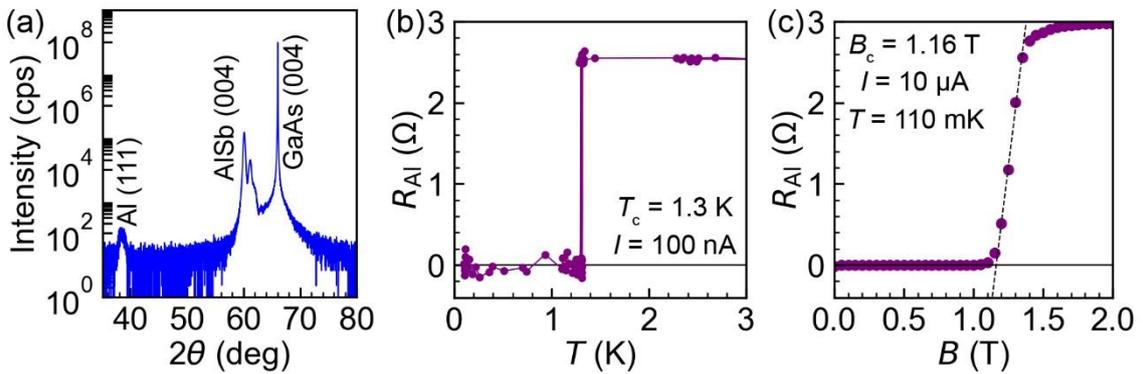

Fig. S1 (a) $\theta$–$2\theta$ X-ray diffraction scan of the sample structure shown in Fig. 2(a). (b) Temperature dependence of the Al resistance measured with a current of 100 nA. (c) Magnetic field dependence of the Al resistance. The magnetic field was applied in-plane along the [$\bar{1}$10] direction of the GaAs (001) substrate. The measurement current was 10 μA, and the temperature was 110 mK.

**Ferromagnetic properties and calculated stray field generated by (Ga,Fe)Sb**

The ferromagnetism of the (Ga,Fe)Sb layer in our sample was evaluated using two techniques, magnetic circular dichroism (MCD) and superconducting quantum interference device (SQUID) magnetometry. Figure S2(a) shows the MCD spectra of (Ga,Fe)Sb measured at 5 K. All spectra obtained under magnetic fields of 1, 0.5, and 0.2 T exhibit identical spectral features at the critical point energies $E_1$ and $E_1 + \Delta_1$ of GaSb, with no additional peaks. This indicates that the ferromagnetism originates intrinsically from the zinc-blende-type (Ga,Fe)Sb rather than from Fe clusters or secondary Fe-containing compounds.

Figure S2(b) presents an Arrott plot derived from Fig. 2(e), which is an effective method for estimating the Curie temperature $T_C$ from magnetic hysteresis loops [S2]. A positive (negative) intercept of the linear fit corresponds to a ferromagnetic (paramagnetic) state. The intercept crosses the origin at 25 K, indicating that $T_C$ of (Ga,Fe)Sb in this sample is 25 K. This value is lower than those reported in previous studies [S3], likely due to non-optimal growth conditions, as the magnetic properties of FMSs grown by low-temperature MBE are highly sensitive to flux ratios, substrate temperature, and other growth parameters.

Figure S2(c) shows an enlarged hysteresis loop of (Ga,Fe)Sb measured by SQUID (also shown in Fig. 2(e)), exhibiting a remanent magnetisation of approximately 15 emu/cm$^3$. Using this magnetisation value, the stray magnetic field generated by the (Ga,Fe)Sb layer was calculated at the center of the InAs layer, as shown in Fig. S2(d), to understand its influence on the Josephson junction behaviour under magnetic fields. The calculation was performed based on Maxwell's equations as described in Ref. [S4]. The (Ga,Fe)Sb layer was modeled with dimensions of 30 μm × 5 μm × 20 nm, corresponding to the size of the central electrode in the Josephson junction device. The magnetisation was assumed to be uniaxial and oriented perpendicular to the plane. The calculated results, shown in Fig. S2(e), indicate that the stray field at the InAs layer generated by (Ga,Fe)Sb is perpendicular to the plane and on the order of 0.01–0.1 mT near zero external field. Even when the flux focusing effect arising from the Meissner response of the superconducting electrodes is taken into account, this stray field alone cannot fully explain the magnetic-field positions of the critical-current peaks observed in the Fraunhofer patterns of our junction devices (Fig. 4 (a) and (b)), as discussed in the next section. Nonetheless, the calculated stray field may offer some indication of how magnetic fields influence the superconducting channel.

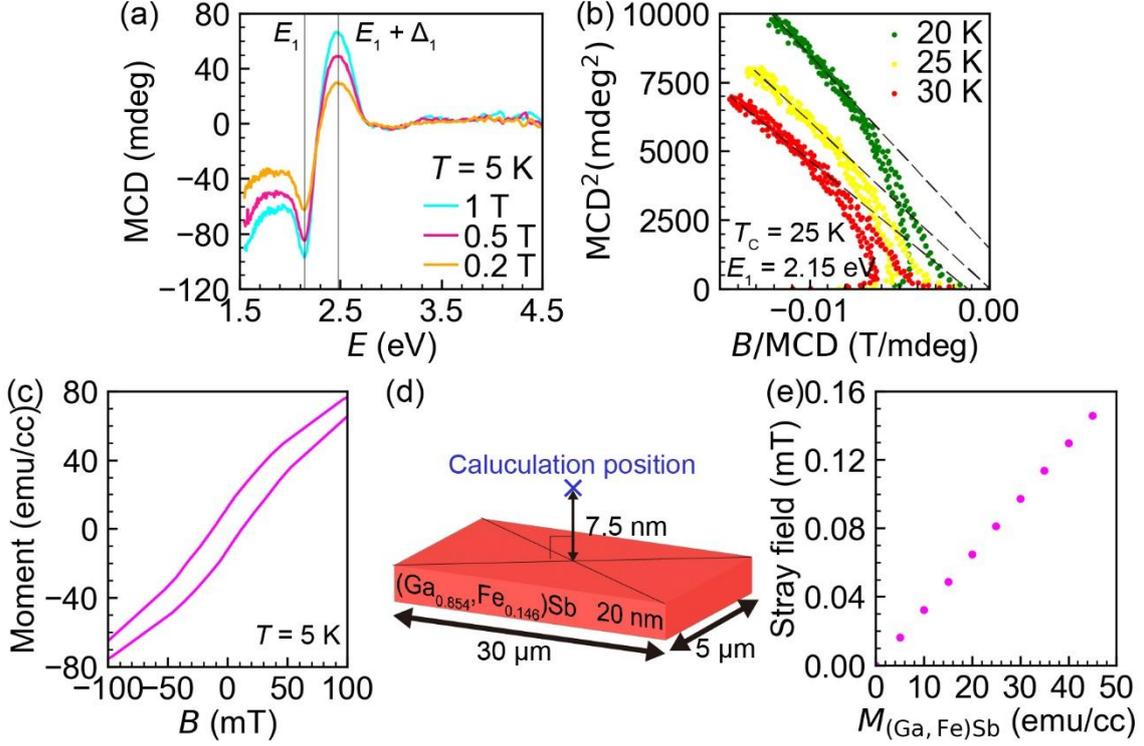

Fig. S2 (a) MCD spectra of (Ga,Fe)Sb measured at 5 K under external magnetic fields of 1, 0.5, and 0.2 T applied perpendicular to the film plane. (b) Arrott plot of (Ga,Fe)Sb obtained at the photon energy $E_1$ = 2.15 eV. (c) Magnetic hysteresis of (Ga,Fe)Sb measured by SQUID at 5 K when a magnetic field is applied perpendicular to the film plane. (d) Schematic geometry used for calculating the stray magnetic field from (Ga,Fe)Sb at the InAs layer. (e) Calculated stray-field magnitude at the position indicated in (d) as a function of the perpendicular magnetisation of the (Ga,Fe)Sb layer.

**Discussion about the magnetic field where the critical current reaches maximum in Fraunhofer patterns**

To further understand the Fraunhofer interference observed in the FMS-based Josephson junctions (JJs), it is necessary to examine the magnetic field at which the maximum critical current occurs, denoted $B_{shift}$ in Fig. 4(c). Notably, clear differences appear between Figs. 4 and 5. In Fig. 4, the ungated junction shows its maximum critical current at a field aligned with the initial field, with a magnitude of ~0.4 mT. In contrast, the gated junction (Fig. 5) exhibits the maximum of $I_c$ at a magnetic field opposite to the initial direction, with a magnitude of ~1 mT. This discrepancy raises an unresolved question of what determines the peak positions in the Fraunhofer patterns. Considering the device structure and the possible presence of edge currents, variations in current distribution—potentially arising from differences in the surface conditions of the

junction—may shift the fields at which the critical current peaks appear. Regarding the magnetic properties of (Ga,Fe)Sb, its coercive field is ~20 mT, which cannot account for the observed peak fields. However, the stray field of (Ga,Fe)Sb at the InAs layer with considering flux focusing effect in JJs is calculated to be on the order of 0.1 mT (Fig. S2(e)), comparable but slightly less than the peak positions (~1 mT) observed in Fig. 5 (a)-(e). To fully understand the magnetic field dependence of the critical current peaks, further investigation is necessary through both experimental studies and numerical simulations.

**Comparison of gate-voltage dependences of the Josephson critical current and Hall-bar carrier density**

Figure S3(a) shows the gate voltage dependence of the critical current $I_c$ in the gated Josephson junction (JJ). The blue data points are extracted from the $V-I$ measurements shown in Fig. 3(f) before applying a magnetic field, while the green points correspond to the maximum $I_c$ values obtained from the Fraunhofer patterns. No $I_c$ data are available between −10 and 0 V from the Fraunhofer measurements due to the measurement procedure; however, we assume that the $I_c$ obtained from the $V-I$ data before applying magnetic field corresponds to the maximum $I_c$ in the Fraunhofer patterns. Indeed, the blue and green data connect smoothly, and these combined data are referred to as peak 1 in Fig. 5(e), (f) in the main text. The junction exhibits a decrease in $I_c$ with increasing the negative gate voltage ($V_{g, JJ}$), with saturation observed above 0 V and around −25 V.

Meanwhile, Fig. S3(b) shows the gate voltage ($V_{g, Hall}$) dependence of the electron carrier density $n$ measured at 5 K using a gated Hall bar device with a conduction path of 200 μm × 50 μm. $n$ decreases with decreasing the gate voltage and nearly saturates above 10 V and around −30 V, following the same overall trend as in Fig. S3(a). Therefore, the decay ratios of $I_c$ and $n$ are compared in Fig. 5(f) within the gate voltage range where both quantities decrease. The difference in the voltage ranges showing the decay of $I_c$ (−25 to 0 V) and $n$ (−30 to 10 V) may arise from structural differences between the two devices, specifically, the gate-oxide thicknesses (40 nm $Al_2O_3$ in the JJ device and 100 nm in the Hall-bar device) which were fabricated by different machines or the presence of Al superconducting electrodes in the junction. Thus, we compare the gate-voltage dependence of $I_c$ and $n$ by normalizing the gate voltage range where both the parameters are effectively tuned. The good agreement between the decay ratios of $I_c$ and $n$ likely suggests that the maximum $I_c$ is approximately proportional to $n$.

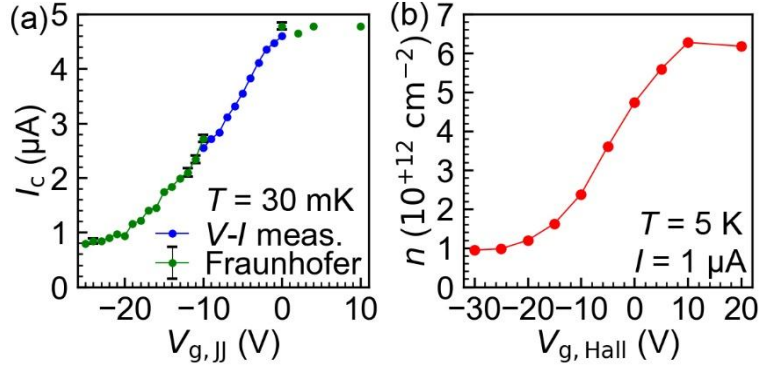

Fig. S3 (a) Gate voltage ($V_g$) dependence of the critical current $I_c$ in the Josephson junction, obtained from V-I measurements before applying an external magnetic field (blue points) and from the maximum values of $I_c$ in the Fraunhofer patterns (green points). (b) Gate voltage dependence of the electron carrier density $n$ measured in a Hall bar device at 5 K with a measurement current of 1 μA.

**Fourier-transformation of the Fraunhofer pattern**

Analyzing a Fraunhofer interference pattern provides key insights into the Josephson junction, such as the effective area where the supercurrent flows and the spatial distribution of the supercurrent density obtained by fast Fourier transformation (FFT). Figure S4(a) shows the magnetic field dependence of the absolute value of the critical current $|I_c|$ in the same junction as in Fig. 4(c). By comparing $I_c^+$ (red points) and $I_c^-$ (blue points), a clearly skewed Fraunhofer pattern is observed, which approximately satisfies $I_c^+(+B) \approx I_c^-(-B)$, $I_c^+(+B) \neq I_c^+(-B)$, and $I_c^+(+B) \neq I_c^-(+B)$.

The solid curve in Fig. S4(a) represents a fit using the conventional Fraunhofer interference expression $I_c(B) = I_0 |\sin(\pi x)/(\pi x)|$, where $x = \Gamma (B-B_{shift}) S / \Phi_0$. Here, $\Gamma$ is a fitting parameter known as the flux focusing coefficient ($\Gamma = 6.2$), and $S$ is the effective junction area penetrated by the magnetic flux, calculated as $S = W \times (L + 2L_{Al})$, where $W$ = 5 μm is the junction width, $L$ = 100 nm is the junction length, and $L_{Al}$ = 190 nm is the magnetic penetration length into the Al electrodes. The value of $L_{Al}$ was obtained following the method described in Ref. [S5], using $\lambda = \lambda_L(0) (1 + \xi/d)^{1/2}$ and $\xi_{thin} = \xi_{bulk}(T_{c,thin}/T_{c,bulk})$, where the parameters are as follows: bulk penetration depth $\lambda_L(0)$ = 16 nm, film thickness $d$ = 10 nm, bulk coherence length $\xi_{bulk}$ = 1.6 μm, experimental $T_{c,thin}$ = 1.3 K, and bulk $T_{c,bulk}$ = 1.2 K [S6][S7]. The large flux focusing coefficient ($\Gamma = 6.2$) indicates that the effective magnetic field in the junction area is approximately six times the external field, likely due to flux concentration arising from the Meissner effect of the Al electrodes and the stray field from the (Ga,Fe)Sb layer.

The supercurrent density distribution in the junction was then calculated by performing a FFT on the data in Fig. S4(a). The coordinate system is defined as follows, $x$ is the current direction, $y$ is the in-plane axis perpendicular to $x$, and $z$ is the perpendicular direction, as illustrated in Fig. 4(e).

The critical current can be expressed as $I_c(\Gamma\beta)$, where $\beta = 2\pi (B-B_{shift}) (L+2L_{Al}) / \Phi_0$. For the FFT analysis, $I_c(\Gamma\beta)$ is treated as a complex function $F(\Gamma\beta)$, satisfying $I_c(\Gamma\beta) = |F(\Gamma\beta)|$. When $I_c(\Gamma\beta)$ is approximately even with respect to $\beta$, it can be written as $F(\Gamma\beta) = I_{c,even}(\Gamma\beta) + iI_{c,odd}(\Gamma\beta)$, where $I_{c,even}$ and $I_{c,odd}$ are the even and odd components of $I_c(\Gamma\beta)$, respectively. The position dependent current density $j(y)$ can then be calculated as $j(y) = \Gamma/(2\pi) \cdot |\int d\beta\, F(\Gamma\beta)\, e^{-i\Gamma\beta y}|$, considering the flux focusing coefficient. The detailed FFT procedure follows Ref. [S8].

The resulting current-density map, shown in Fig. 4(d), reveals that the $I_c^+$ and $I_c^-$ components in Fig. S4(a) are concentrated near the junction edges, meaningfully larger than the FFT result derived from the ideal Fraunhofer fit. This suggests the presence of conduction channels localised at the two edges of InAs in the junction. The fitted current-density distribution shows a slightly reduced supercurrent density in the junction center, likely due to the limited magnetic field range used in the measurement to avoid flux jumps during field sweeps.

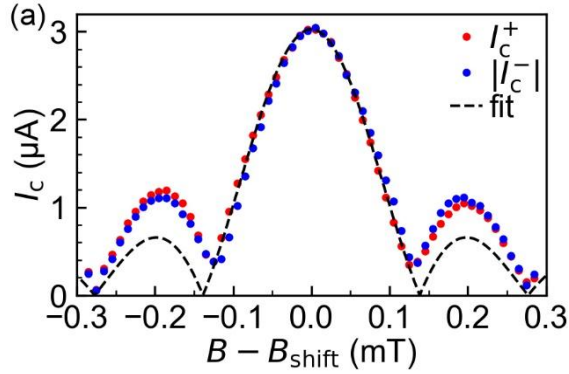

Fig. S4 (a) Magnetic field dependence of the absolute critical current for opposite current directions through the Josephson junction device. The dashed line represents a fit to both $I_c^+$ and $I_c^-$ using the conventional Fraunhofer interference expression.

**In-plane magnetic field dependence of Fraunhofer patterns**

We also investigated the dependence of the Fraunhofer patterns on in-plane magnetic fields in the gated Josephson junction to check whether it is a π-junction. Reentrant superconductivity could appear if the junction were to exhibit π-junction behaviour [S9]-[S12]. For clarity in defining the directions of the applied magnetic fields, a schematic of the gated Josephson junction device and the coordinate axes is shown again in Fig. S5(a). Figure S5(b) shows the critical current $I_c$ as a function of the in-plane magnetic field (parallel to the current) and perpendicular magnetic field, while Fig. S5(d) presents $I_c$ as a function of the in-plane magnetic field (perpendicular to the current) and perpendicular magnetic field. To clearly visualise the maximum $I_c$, the normalised $I_c$ values for each in-plane field are represented in Figs. S5(c) and S5(e). The perpendicular magnetic field corresponding to the $I_c$ maximum is not stable but fluctuates with the in-plane field, rather than exhibiting clear reentrant superconductivity at a fixed perpendicular field. The fluctuation of the $I_c$ maximum is not likely due to field misalignment, because the fluctuations are not linear with applied field, and is not likely due to measurement instability or noise; variations in magnetic field taking $I_c$ maximum is only 0.2mT, which is much smaller than the observed fluctuations, with multiple scans (5 times) at one in-plane magnetic field. Although the origin of these fluctuations remains unclear, this behaviour possibly represents a feature of Josephson junctions based on FMS heterostructures.

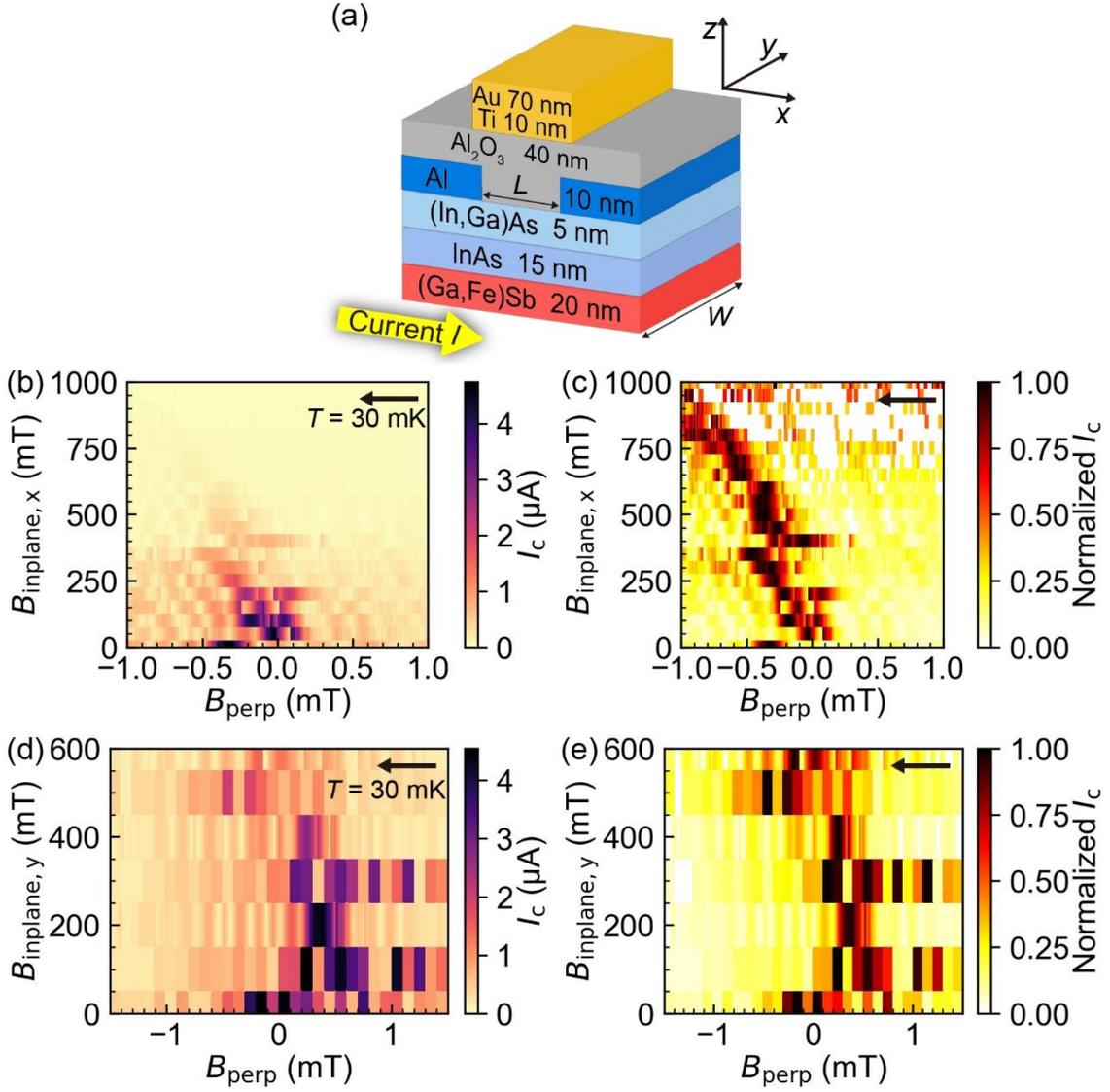

Fig. S5 (a) Schematic illustration of a gated Josephson junction examined in this study and the definition of coordinate axes *x*, *y*, and *z*. (b) Dependence of the critical current $I_c$ on the in-plane (*x*) and perpendicular (*z*) magnetic fields. (c) Normalised $I_c$ for each in-plane magnetic field $B_x$ from (b). (d) Dependence of $I_c$ on the in-plane (*y*) and perpendicular (*z*) magnetic fields. (e) Normalised $I_c$ for each in-plane magnetic field $B_y$ from (d). Black arrows indicate the sweep directions of the perpendicular magnetic field.